# Microwave vortex beam launcher design

Nedime Pelin M. H. Salem [1*], Edip Niver [1], Mohamed A. Salem [2]

[1] Electrical and Computer Engineering Department, New Jersey Institute of Technology, Newark, NJ, USA
[2] Engineering Science Department, Sonoma State University, Rohnert Park, CA, USA
[*] pelin.salem@njit.edu

**Abstract:** A novel design for a vectorial vortex beam launcher in the microwave regime is devised. The beam is formed by launching a single guided transverse electric (TE) mode of a metallic circular waveguide into free-space. Excitation is achieved by the mean of an inserted coaxial loop antenna. Modal expansion coefficients are computed, and the resulting electric and magnetic fields are determined. The effect of the antenna location inside the waveguide on its effective input impedance is modelled using transmission-line relations and location for optimal matching is established. The analytical results are confirmed using multi-level fast multipole method full-wave simulations.

## 1. Introduction

Vector vortex beams are monochromatic electromagnetic wave fields carrying spin angular momentum and orbital angular momentum (OAM). Spin angular momentum is associated with the polarisation of the field, whereas OAM yields an azimuthal field dependence of the form $\exp(\pm jq\phi)$, where $\phi$ is the azimuthal angle, and $q$ is an integer designating the helicity order, which is also known as the topological charge of the vortex beam [1, 2]. Vortex beams owe their names to the characteristic on-axis phase singularity and amplitude null [3].

Interest in OAM-carrying beams spans a variety of applications, such as imaging [4, 5], quantum applications [6–8], electron microscopy [9], and astronomy [10]. Particular interest is taken in employing vortex beams in optical and wireless communication, because of the orthogonality of OAM states. This orthogonality property is exploited as a viable diversity in the form of OAM multiplexing, to increase the communication capacity over optical fibre and wireless channels [11, 12].

Various approaches for vortex beam generation in the microwave regime are presented in the literature. An approach is implemented in [13] by connecting eight Vivaldi antenna elements sequentially and folding them into a hollow cylinder to configure a circular Vivaldi antenna array. An alternative method for launching OAM carrying vortex beams employing a circular leaky wave antenna is implemented in [14]. In [15], a Cassegrain reflector antenna fed by a $2 \times 2$ matrix composed of four open-ended rectangular waveguides is implemented for simultaneous generation of three OAM modes ($q = 0, \pm1$) as part of a space diversity scheme for wireless communications. A multilayer amplitude-phase-modulated surfaces transmit-array based on four-layer conformal square-loop elements is proposed in [16] for generating a second-order Bessel vortex beam with OAM in the radio-frequency domain. In [17], OAM states are generated by exciting a combination of circular waveguide modes in quadrature and propagating the field through a modal-transformation waveguide section. The OAM-carrying field is radiated in free-space through a conical horn antenna.

In this study, we devise a new method to generate vector vortex beams in the microwave regime based on waveguide modes, where the vortex beam is set to be the aperture field at the open end of a metallic circular waveguide section. For simplicity and without loss of generality, a transverse electric (TE) beam with a truncated Bessel profile is considered. The aperture field is formed by the propagating field of the $TE_{q1}$ mode of the waveguide, where $q$ is the topological charge of vortex beam to be launched. Excitation is provided by means of a single circular loop antenna inserted coaxially inside the waveguide. In the proposed design, the waveguide housing the large loop antenna is shown to be advantageous in terms of matching since the antenna input impedance depends on the antenna location inside the waveguide. A simplified analytical expression of the input impedance of the loop antenna inside the waveguide is derived based on transmission-line (TL) theory and verified again by the multi-level fast multipole method (MLFMM) full-wave simulation. Full-wave simulation is also employed to verify the function of the proposed design. This proposed method takes inspiration from previous work on zero-order Bessel beam generation in the microwave regime [18] but accounts for the practical aspects, namely, that the current on the antenna is not constant and that the antenna input impedance is matched to the source. In addition, the proposed method permits reducing the number of required antennas to a single antenna.

In what follows, the structure of the launcher is detailed in Section 2. Modal analysis of the field excited by the loop antenna is detailed in Section 3. The analysis establishes the necessary conditions for single $TE_{q1}$ mode excitation. An analytical model of the antenna input impedance as a function of its location inside the waveguide is developed based on physical considerations using TL theory is given in Section 4. In Section 5, the model of the proposed launcher is applied for launching a truncated Bessel beam carrying OAM with topological charge $q = 4$ and is verified using MLFMM full-wave simulation. Conclusions of the study are given in Section 6.

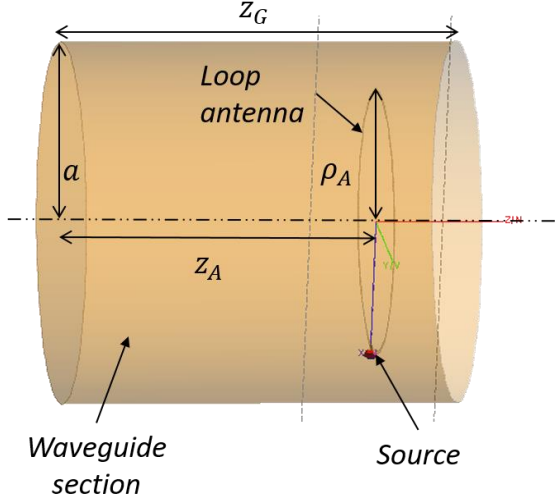

*Fig. 1: A schematic of the proposed vortex beam launcher structure.*

## 2. Vortex beam launcher design

The structure of the proposed launcher consists of two main components: (i) a thin sheet loop antenna connected to a voltage source with 50 Ω impedance and (ii) a finite circular waveguide section with one closed end. The antenna and waveguide are coaxially aligned along the $z$–axis, as illustrated in Fig. 1. The radius of the waveguide section $a$ is chosen such that

$$p'_{q1} < ak_0 < p_{q1}, \quad (1)$$

where $k_0 = 2\pi/\lambda_0$ is the free-space wavenumber, and $p_{q1}$ and $p'_{q1}$ are the first roots of the $q$th order Bessel function of the first kind, and its derivative, respectively. This particular choice of $a$ ensures that the $\text{TE}_{q1}$ mode is propagating, while the next mode with same azimuthal dependence, the transverse magnetic $\text{TM}_{q1}$, is in cut-off. The radius of the loop antenna $\rho_A$ is chosen such that the circumference length corresponds to $q\lambda_0$, where $\lambda_0$ is the operation wavelength in free-space. This choice of circumference length together with the choice of the waveguide radius ensure that the antenna excites the $\text{TE}_{q1}$ mode of the circular waveguide section, which is launched into free-space through the open end. The profile of the launched vortex beam thus has a 'doughnut' shape, since the radial dependence of the $\text{TE}_{q1}$ field corresponds to $q$th order Bessel function truncated at its first zero. The distance of the antenna from the closed end $z_A$ is determined by the choice of $q$ to maximise the power coupling efficiency to the $\text{TE}_{q1}$ mode. The length of the waveguide section $z_G$ is shown to have little effect on the overall performance of the launcher, as long as, roughly speaking, $z_G \geq 0.75\lambda_{g,q1}$, where $\lambda_{g,q1} = 2\pi/\beta_{q1}^{\text{TE}}$ and $\beta_{q1}^{\text{TE}}$ is the propagation constant of the $\text{TE}_{q1}$ mode.

## 3. Single propagating mode excitation

The electromagnetic field excited by a time-harmonic source placed inside a circular waveguide radiates, if propagating, or decays, if evanescent, in the direction away from the source region. The excited field due to a current source $J(\rho, \phi, z)$ may be expressed as superposition of the circular waveguide eigenmodes, such as [19]

$$\begin{aligned} \boldsymbol{E}^+(\rho,\phi,z) &= \sum_\nu a_\nu \boldsymbol{E}_\nu^+(\rho,\phi,z), \\ \boldsymbol{H}^+(\rho,\phi,z) &= \sum_\nu a_\nu \boldsymbol{H}_\nu^+(\rho,\phi,z), \\ \boldsymbol{E}^-(\rho,\phi,z) &= \sum_\nu b_\nu \boldsymbol{E}_\nu^-(\rho,\phi,z), \\ \boldsymbol{H}^-(\rho,\phi,z) &= \sum_\nu b_\nu \boldsymbol{H}_\nu^-(\rho,\phi,z), \end{aligned} \quad (2)$$

where $\boldsymbol{E}$ and $\boldsymbol{H}$ are the electric and magnetic fields, respectively, $(\rho, \phi, z)$ are the circular cylindrical coordinate variables, superscripts $\pm$ designate fields radiating or decaying in the $\pm z$–direction, respectively, $\nu$ is an index for a waveguide eigenmode, and $a_\nu$ and $b_\nu$ are modal excitation coefficients. The harmonic time-dependence is assumed to be $\exp(j\omega t)$, with $\omega$ the angular frequency, and is omitted throughout.

Using the Lorentz reciprocity principle on the current source with its field and the eigenmode $\sigma$ yields the relation

$$\oiint_S (\boldsymbol{E}_\sigma^\pm \times \boldsymbol{H}^\pm - \boldsymbol{E}^\pm \times \boldsymbol{H}_\sigma^\pm) \cdot \hat{\boldsymbol{n}} dS = -\iiint_V \boldsymbol{J} \cdot \boldsymbol{E}_\sigma^\pm dV, \quad (3)$$

where $V$ is a volume enclosing the source region, and $S$ its surface with $\hat{\boldsymbol{n}}$ the unit normal to $S$.

Substituting (2) into (3) and applying the mode orthogonality relation, it is shown that [19]

$$\begin{aligned} a_\nu &= -\frac{1}{2}\iiint_V \boldsymbol{J} \cdot \boldsymbol{E}_\nu^- dV, \\ b_\nu &= -\frac{1}{2}\iiint_V \boldsymbol{J} \cdot \boldsymbol{E}_\nu^+ dV. \end{aligned} \quad (4)$$

For the proposed launcher design, the current density on the loop antenna may be modelled as

$$\boldsymbol{J}(\rho,\phi,z) = A e^{-jq\phi} \frac{\delta(\rho-\rho_A)}{\rho}\delta(z-z_A)\hat{\boldsymbol{\phi}}, \quad (5)$$

where $A$ is an arbitrary excitation amplitude, $\delta(\cdot)$ is the Dirac delta function, and $\hat{\boldsymbol{\phi}}$ is the unit vector in the $\phi$–direction. Substituting (5) and the $\text{TE}_{nm}$ and $\text{TM}_{nm}$ electric field expressions from [20] into (4) yields the following mode excitation coefficients:

$$\begin{aligned} a_{nm}^{\text{TE}} &= -A\pi e^{\Gamma_{nm}^{\text{TE}} z_A} \frac{j\omega\mu_0}{k_{c,nm}} J'_n(k_{c,nm}\rho_A)\delta_{nq}, \\ a_{nm}^{\text{TM}} &= -A\pi e^{\Gamma_{nm}^{\text{TM}} z_A} \frac{j\Gamma_{nm}^{\text{TM}} n}{k_{c,nm}^2 \rho_A} J_n(k_{c,nm}\rho_A)\delta_{nq}, \\ b_{nm}^{\text{TE}} &= -A\pi e^{-\Gamma_{nm}^{\text{TE}} z_A} \frac{j\omega\mu_0}{k_{c,nm}} J'_n(k_{c,nm}\rho_A)\delta_{nq}, \\ b_{nm}^{\text{TM}} &= A\pi e^{-\Gamma_{nm}^{\text{TM}} z_A} \frac{j\Gamma_{nm}^{\text{TM}} n}{k_{c,nm}^2 \rho_A} J_n(k_{c,nm}\rho_A)\delta_{nq}, \end{aligned} \quad (6)$$

where $\mu_0$ is the free-space magnetic permeability, $J_n(\cdot)$ and $J'_n(\cdot)$ are the $n$–th order Bessel function of the first kind and its derivative, respectively, $k_{c,nm}^{\text{TE}} = p'_{nm}/a$ and $k_{c,nm}^{\text{TM}} = p_{nm}/a$ are the TE and TM mode transverse wavenumbers, respectively, $\Gamma_{nm}^2 = k_0^2 - k_{c_{nm}}^2$, and $\delta_{nq}$ is the Kronecker delta function. It is thus evident that only $\text{TE}_{qm}$ and $\text{TM}_{qm}$ modes may be excited by the loop antenna current. Furthermore, the choice of $a$ according to (1) ensures that TE$_{q1}$ is the only propagating mode inside the waveguide, as required by the design.

## 4. Input impedance model and matching

The necessary conditions for single TE$_{q1}$ mode operation determine the value of $\rho_A$ and set upper and lower bounds for the value of $a$. The remaining design parameters, namely, $z_A$ and $z_G$ are determined by the necessary conditions to optimally match the loop antenna to the voltage source with 50 Ω internal impedance.

Large loop antennas are known to be difficult to match to common voltage sources due to their large input impedance and specifically their large inductive reactance [21]. However, by placing the loop antenna inside the waveguide section, the antenna input impedance becomes dependent on where the antenna is placed relative to the end of the waveguide section. This additional degree of freedom allows the antenna to be optimally matched to the voltage source. In what follows, and instead of using an exact method to determine the antenna input impedance as a function of its location, such as the method suggested in [22], a phenomenological approach is employed based on TL theory that takes into account the correct field behaviour. The TL approach yields a simpler mathematical expression of $Z_{in}$, which is faster to evaluate and provides greater physical insight without sacrificing accuracy.

In this proposed input impedance model, the effect of launcher structure is divided into distinct physical contributions each modelled as a terminated TL section. This approach is suitable for the proposed design because the TE$_{q1}$ mode is the only propagating mode inside the waveguide section. Accordingly, the propagation constant of the TL is

$$\beta_g = \beta_{q1}^{\text{TE}} = \sqrt{k_0^2 - \left(k_{c,q1}^{\text{TE}}\right)^2},$$

and the TL characteristic impedance is $Z_g = (k_0/\beta_g)Z_0$, where $Z_0 \approx 377$ Ω is the free-space impedance.

The input impedance of the loop antenna inside the launcher may be, to a first-order approximation, modelled as a parallel combination of
(i) the free-space loop antenna input impedance $Z_A$,
(ii) a short-circuited TL of length $z_A$ with impedance $Z_{\text{CE}}$ at the antenna terminal, where
$$Z_{\text{CE}} = jZ_g \tan(\beta_g z_A),$$
which models the effect of the closed end of the waveguide section,
(iii) a $Z_0$ terminated TL of length $z_o = z_G - z_A$ with impedance $Z_{\text{OE}}$ at the antenna terminal, where
$$Z_{\text{OE}} = \frac{Z_g[Z_0 + jZ_g \tan(\beta_g z_o)]}{[Z_g + jZ_0 \tan(\beta_g z_o)]},$$
which models the effect of the open end of the waveguide section,
(iv) a TL terminated by $Z_0$ in series with $Z_{sc} = jZ_g\tan(\beta_g z_G)$, of length $z_o$ with impedance $Z_R$ at the antenna terminal, where
$$Z_R = \frac{Z_g[Z_0 + Z_{sc} + jZ_g \tan(\beta_g z_o)]}{[Z_g + j[Z_0 + Z_{sc}] \tan(\beta_g z_o)]},$$
which models the effect of reflection from the open end of the field reflected from the closed end.

In addition to the previous contributions, the coupling between the loop antenna and the waveguide may be modelled by a series impedance $Z_{\text{CC}} = -2jX_A \sin(\beta_g z_o)$, where $X_A$ is the reactance of the loop antenna in free space. The input impedance is thus given by

$$Z_{in} = Z_{\text{CC}} + \left(\frac{1}{Z_A} + \frac{1}{Z_{\text{CE}}} + \frac{1}{Z_{\text{OE}}} + \frac{1}{Z_R}\right)^{-1} \quad (7)$$

The model shows that if $z_A = \lambda_{g,q1}/2$, the input resistance drops to zero. This suggests that the optimal choice of $z_A$ is in the vicinity of $\lambda_{g,q1}/2$ such that the resistive part of the input impedance is 50 Ω. Since there are two possible $z_A$ choices resulting in Re$\{Z_A\} = 50$ Ω, preference is given to the one with inductive reactance, because it is more practical to match using a capacitive load. From the expression of $Z_{\text{CE}}$, the optimal placement is thus expected to exist at $z_A > \lambda_{g,q1}/2$. Moreover, it could be seen that the effect of $z_G$ does not significantly affect the overall behaviour of $Z_{in}$ as long as it is chosen such that $z_G \geq 3\lambda_{g,q1}/4$ to avoid affecting the optimal matching position. Nevertheless, it should be noted that (7) is not accurate in estimating the input impedance, specifically its reactance, away from the ends of the waveguide section, which suggests that additional coupling effects should be taken into consideration. This shortcoming, however, does not affect its validity in estimating $Z_A$ near the optimal matching position. Overall, the model provides sound guidelines to find the optimal antenna location for the best matching performance, as shall be shown in the following section.

## 5. Fourth-order Bessel beam launcher

The proposed launcher design is next applied for launching a truncated microwave Bessel beam in the X-band carrying OAM with topological charge $q = 4$. The operation frequency is chosen to be 9 GHz. The structure excites the TE$_{41}$ mode of the circular waveguide section, which is launched into free space through the open end. The radius of the waveguide is chosen as $a = 30$ mm, which sets the cut-off frequency of the TE$_{41}$ mode at 8.47 GHz, ensuring operation above cut-off. The structure operates at single mode, since the next mode with the same angular dependence, TM$_{41}$, has its cut-off frequency at 12.09 GHz, which is above the operation frequency.

The operation of the designed launcher is verified by MLFMM full-wave simulation performed using the default settings of Altair Hyperworks FEKO simulation software [23]. The loop antenna is constructed as a strip loop using perfect electric conductor (PEC) material. The antenna radius is set to $\rho_A = 21.22$ mm, corresponding to a circumference of $4\lambda_0$ with $\lambda_0 = 33.31$ mm and the width of the strip is set to 0.5 mm. An edge port with a voltage source of 1 V and 50 Ω impedance is connected to the antenna feed line. The waveguide section is constructed using PEC material, with its radius set to $a = 30$ mm, its length set to $z_G = 64$ mm, and its closed end at $z = 0$. The simulation frequency is set to 9.0 GHz. The launcher structure geometry is meshed using fine global mesh setting, and re-mesh operation is performed after every modification done to the design.

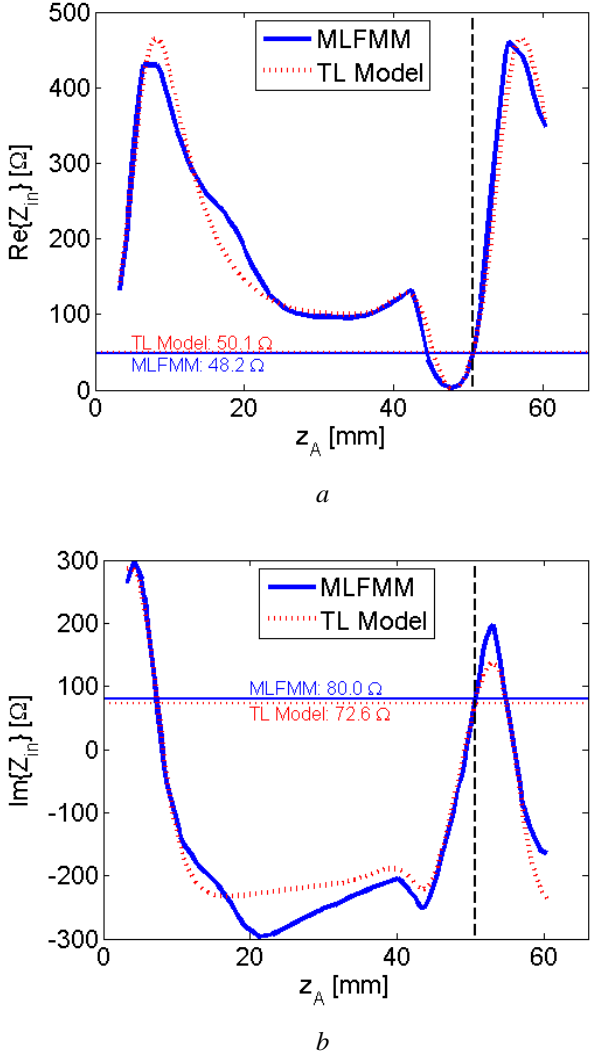

*Fig. 2: Comparison between full-wave simulated and TL modelled (a) real part, and (b) imaginary part of the loop antenna input impedance versus its position inside a launcher structure with $z_G = 64$ mm. The radius of the antenna is $\rho_A = 21.22$ mm and the antenna excites the $TE_{q1}$ mode at operation frequency of 9 GHz. The vertical lines at $z_A = 50.5$ mm indicate the optimal antenna position for matching.*

The position of the loop antenna $z_A$ is varied from $z = 2.12$ mm to $z = 61.51$ mm in 28 steps and $Z_{in}$ is measured. Fig. 2 plots the real and imaginary parts of $Z_{in}$ as obtained by simulation and by (7). The figure shows very good match between the simulation and the model for the real part with maximum deviation of $69.44\,\Omega$ at $z_A = 54.90$ mm and a root mean square (rms) deviation of $25.78 \pm 18.12\,\Omega$ over the entire test range. A good match for the imaginary part is observed when the antenna is close to the ends of the waveguide with a maximum deviation of $68.33\,\Omega$ at $z_A = 21.27$ mm and an rms deviation of $34.53 \pm 18.90\,\Omega$ over the entire range. The TL model gives lower estimate for the reactance value away from the waveguide ends, which suggests that additional coupling, potentially due to more energy stored in evanescent modes, has to be taken into consideration in this region.

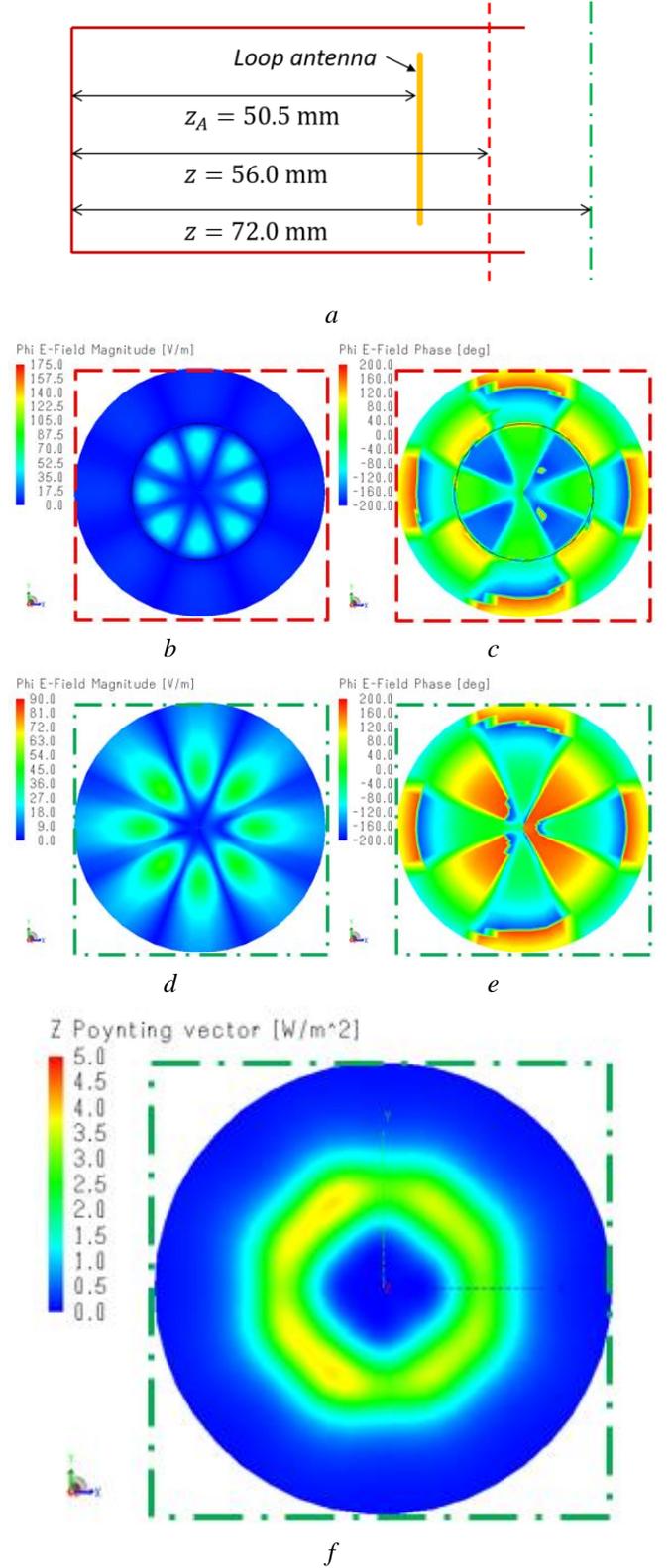

*Fig. 3: Plots of the electric field at different $z$–planes as shown in the schematic in (a). The (b) magnitude and (c) phase of $E_\phi$ at $z = 56$ mm inside the launcher, and the corresponding (d) magnitude and (e) phase at $z = 72$ mm in free-space. The phase plots show a helicity pertaining to $q = 4$ topological charge. The $z$–directed component of the Poynting vector is plotted at $z = 72$ mm in free-space in (f) and exhibits the expected 'doughnut-shape' profile.*

For optimal matching condition, the antenna is placed at $z_A = 50.5$ mm, where both (7) and simulation estimate that $Z_{in} = 50 + j80\,\Omega$. To match the antenna, a capacitor with capacitance of 221 fF is placed in series with the source. In practical implementation, this capacitance value can be obtained using one standard 0.22 pF capacitor or two standard 0.11 pF capacitors in parallel.

The electromagnetic performance of the proposed design is measured through full-wave simulation. Figs. 3*b* and *c* plot the magnitude and phase, respectively, of the azimuthal component of the electric field $E_\phi$ at $z = 56$ mm inside the launcher structure. Figs. 3*d* and *e* plot the corresponding values for the launched beam at $z = 72$ mm in free-space outside the launcher. In both cases, the phase exhibits the expected helicity pertaining to the $q = 4$ topological charge, while the magnitude displays a hollow centre due to the phase discontinuity. The $z$–directed component of the Poynting vector, plotted in Fig. 3*f*, clearly shows the 'doughnut-shape' beam profile due to truncation of the Bessel function at its first zero. The full-wave simulation thus verifies that the proposed launcher is indeed capable of launching truncated vector Bessel beams carrying OAM with topological charge $q = 4$.

## 6. Conclusion

In this work, a novel design for a vectorial vortex beam launcher in the microwave regime is proposed and analysed. The vortex beam is formed by launching a single guided $TE_{q1}$ mode of a metallic circular waveguide section into free space. The mode is excited by a single-loop antenna positioned coaxially inside the waveguide section. Modal analysis is carried out to verify the single mode operation of the proposed design. The effect of the position of the antenna inside the waveguide section on its input impedance is modelled using TL relations, and the optimal position for impedance matching is established. Full-wave simulation for an implementation of the launcher design to generate a truncated Bessel beam at 9 GHz carrying OAM with topological charge $q = 4$ is performed. The simulation showed good agreement with the TL impedance model, which is employed to estimate the antenna position for optimal matching. Furthermore, the electromagnetic performance of the launcher is verified showing the expected phase helicity and 'doughnut-shaped' beam profile outside the waveguide section.

The proposed design employing a circular waveguide section is demonstrated to be advantageous when employing large loop antennas in radiating systems. The traditional challenge of matching a large loop antenna to an external source may be overcome by encasing the antenna in a waveguide. In addition, by using a metallic waveguide and a simple loop antenna, the launcher may thus be more suitable for high power applications compared to other designs that rely on dielectric materials or integrated electronics.

Future work may include modelling a coaxial feed line for the antenna, generalising the launcher design to accommodate multimode excitation for beam-shaping purposes, the possibility of benefiting from artificial dielectric filling and improving the TL model of the input impedance.